# Tunable Microwave Magnetic Field Detection based on Rabi Resonance with a Single Cesium-Rubidium Hybrid Vapor Cell

Fuyu Sun,[1] Zhiyuan Jiang,[2] Jifeng Qu,[2] Zhenfei Song,[2] Jie Ma,[1] Dong Hou,[3] and Xiaochi Liu[2,a)]

[1]*Key Laboratory of Time and Frequency Primary Standards, National Time Service Center, Chinese Academy of Science, Xi'an, Shaanxi, 710600, China*

[2]*National Institute of Metrology, Beijing, 100029, China*

[3]*Time & Frequency Research Center, School of Automation Engineering, University of Electronic Science and Technology of China, Chengdu, 611731, China*

We experimentally investigated Rabi resonance-based continuously frequency-tunable microwave (MW) magnetic field detection using a single hybrid vapor cell filled with cesium and rubidium atoms. The multispecies atomic systems, with their tunable abilities in transition frequencies, enabled this atomic sensing head to cover a broader detectable MW field scope compared to the use of a single metal atom. Here, we demonstrated the simultaneous observation of atomic Rabi resonance signals with $^{85}$Rb, $^{87}$Rb, and $^{133}$Cs in the same vapor cell. Using an experimentally feasible static magnetic field (DC field) below 500 Gauss, we realized a MW magnetic field strength detection with bandwidths of 4.8 GHz around 8.1 GHz. The use of these three atomic systems confined in a single vapor cell also enabled the establishment of an identical MW field with the help of DC field, allowing us to perform a perfect comparison for different applications that require the same electromagnetic environment. The results may be useful for the realization and application of many atomic detectors based on different physical principles.

Recently, various atom-based microwave (MW) field detection techniques have been developed, such as MW electrometry using a Rydberg atom[1-4], MW magnetometry based on Rabi oscillations[5-8], and MW magnetic field sensors or standards based on Rabi resonance[9-11]. Atomic-based MW detection techniques are applied widely in many fields, including inertial navigation, detection of explosions, and medical imaging of the heart and brain[12-17]. In these techniques, the strength of the MW field could be translated into a Rabi frequency via atomic constants compared to conventional MW field detection techniques. Therefore, atomic-based MW field sensors have the advantage of a high measurement accuracy and are SI-traceable[18-21]. However, these sensing techniques are generally based on single atomic specie. In the other high-accuracy measurements, multispecies atoms are becoming an ideal system for improving performance and analyzing systematic effects[22-24]. In this

---

a) Author to whom correspondence should be addressed. Electronic mail: liuxch@nim.ac.cn

paper, we report the principle of a continuous frequency tunable MW magnetic field detection technique based on Rabi resonance in a single Cs-Rb hybrid vapor cell with a feasible DC field in a laboratory environment. It is the first time to apply the multispecies atomic system on the MW magnetic field sensing based Rabi resonance to our knowledge. Detectable, continuously tunable MW frequency bandwidths of 4.8 GHz were demonstrated, which partially covered the S band (2 ~ 4 GHz), C band (4 ~ 8 GHz), and X band (8 ~ 12 GHz) of a MW field using $^{85}$Rb, $^{87}$Rb, and $^{133}$Cs, respectively, as the MW field sensing head. The development of this technique highlights the potential for developing atomic-based and SI-traceable magnetometry operating at any frequency.

In our early work[11,25], we reported a Rabi resonance based-MW detection technique at a fixed frequency of 9.19 GHz with a Cs-buffer gas vapor cell, and demonstrated its application to evaluate MW field strength inside a cavity. The general principle of detecting an unknown strength MW field with Rabi resonance is based on a theoretical model developed by J. Camparo et al.[19-21]. The amplitude of the Rabi resonance signal $P_\beta^0$ can be described by Eq. (1) below when the given Zeeman sublevels of alkali atoms resonate with a phase modulated MW field.

$$P_\beta^0 = \frac{1}{4} \frac{m^2 \omega_m \Omega^2 \gamma_2}{[\gamma_2^2 + \Delta^2 + (\gamma_2/\gamma_1)\Omega^2]\sqrt{(\Omega^2 - 4\omega_m^2)^2 + 4\gamma_1^2 \omega_m^2}} \quad (1)$$

where $\gamma_1$ and $\gamma_2$ are the longitudinal and transverse relaxation rate, respectively, $\Delta$ is the average MW field-atom detuning, $\Omega$ is the Rabi frequency, m represents the amplitude of the modulation signal, and $\omega_m$ is the phase modulation frequency.

A Rabi resonance signal lineshape could be plotted by scanning the phase modulation frequency $\omega_m$ of a MW field. The Rabi resonance signal achieves a maximum of $\frac{1}{8}\frac{m^2 \Omega^2}{\Omega^2 + \gamma_1 \gamma_2}$ when the phase modulation frequency $\omega_m$ equals exactly half of the Rabi frequency of the MW field $\Omega$. Then, the strength of the unknown MW magnetic field can be derived by the measured Rabi frequency of the MW field. Thus, Eq. (1) associates the measurable phase modulation frequency with the strength of the MW magnetic field.

This MW magnetic field detection technique has been successfully explored for different applications, including MW magnetic field stabilization[19], MW magnetometry[9-11,25], characterizations of dielectric material[26] and as an SI-traceable MW power standard[27]. However, the sensing frequency of this MW field detection technique is fixed at the frequency of the Zeeman sublevel transition of alkali atoms, which is 6.83 and 9.19 GHz for [87]Rb and [133]Cs atoms, respectively.

Nevertheless, the Zeeman sublevel transitions are sensitive to DC field variation. Consequently, we could finely tune the frequencies of Zeeman transitions by applying a DC field $H_{DC}$ and then extend the sensing frequency range of MW magnetic field detection based on Rabi resonance. Moreover, our previous experimental setup was developed with a single species of alkali atoms (Cs or Rb) vapor cell. The tunable range of the MW field sensing frequency is still limited. In this study, we developed an experimental setup based on a hybrid vapor cell filled with both Cs and Rb atoms, where Rabi resonance signals were generated from both Cs and Rb, therefore, the sensing frequency range of the MW magnetic field was extended further within a relatively low DC field. Essentially, our technique is applicable to a larger MW field-sensing frequency range because it uses a vapor cell containing more species of atoms and/or applies a stronger DC field.

The energy levels of [133]Cs are shown in Fig. 1 (a). The frequency difference between the hyperfine states of F = 3 and F = 4 was 9.19 GHz when $H_{DC} = 0$. When the DC field increased, the hyperfine Zeeman sublevels $m_F$ diverged from the DC field. There are 15 Zeeman sublevel transitions for [133]Cs atoms, with seven $\pi$ transitions where $\Delta m_F = 0$ and eight $\sigma$ transitions where $\Delta m_F = \pm 1$. For [87]Rb, and the hyperfine structure is similar. There are seven Zeeman sublevels, which are separated by 6.83 GHz according to the value of F (Fig. 1 (b)). Thus, the Zeeman transitions of [87]Rb consist of three $\pi$ transitions and four $\sigma$ transitions. For [85]Rb, the hyperfine splitting is 3.035 GHz, and there are five $\pi$ transitions and six $\sigma$ transitions (Fig. 1 (c)).

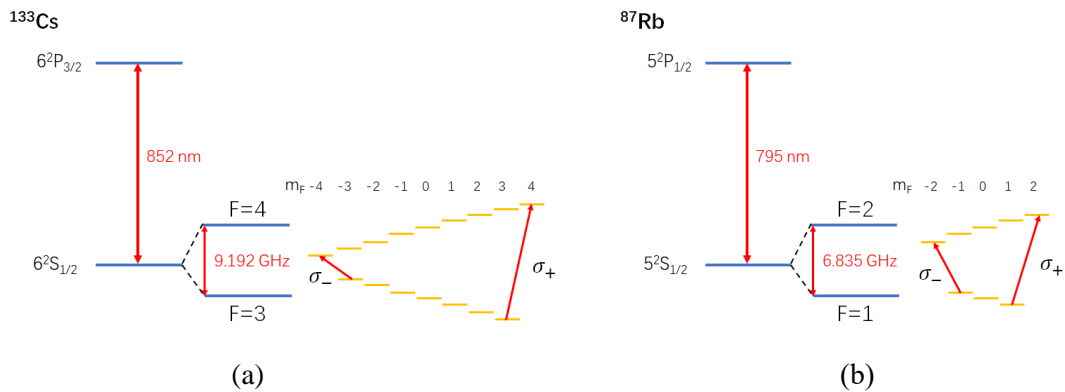

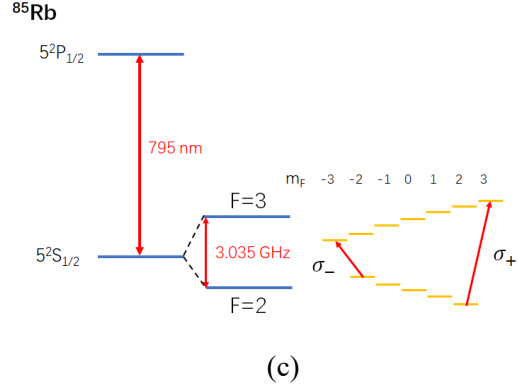

(c)

FIG. 1. Energy levels of $^{133}$Cs (a), $^{87}$Rb (b), and $^{85}$Rb(c).

As the DC field increased, the end Zeeman level $\sigma_+$ transitions $|F = 3, m_F = +3\rangle$—$|F = 4, m_F = +4\rangle$ for Cs and $|F = 1, m_F = +1\rangle$—$|F = 2, m_F = +2\rangle$ $^{87}$Rb were shifted to a higher frequency range, and inversely, the end $\sigma_-$ transitions $|F = 3, m_F = -3\rangle$—$|F = 4, m_F = -4\rangle$ for Cs and $|F = 1, m_F = -1\rangle$—$|F = 2, m_F = -2\rangle$ for $^{87}$Rb were shifted to a lower frequency range. Because the hyperfine splitting frequency in the null DC field of $^{133}$Cs was 9.19 GHz, which is higher than the 6.8 GHz for $^{87}$Rb, the end $\sigma_-$ transition frequency of Cs would be overlapped with the end $\sigma_+$ transition frequency of $^{87}$Rb when a sufficiently strong DC field was applied. Therefore, a continuously tunable frequency range from the end $\sigma_-$ transition of $^{87}$Rb to the end $\sigma_+$ transition of $^{133}$Cs could be realized. Moreover, if $^{85}$Rb atoms were also considered, where the hyperfine splitting frequency was 3.04 GHz, the tunable frequency range could be extended to lower than 2 GHz, with a DC field of ~ 860 Gauss. Figure 2 shows the numerical calculation of the end Zeeman transition frequency as a function of the DC field for $^{85}$Rb, $^{87}$Rb, and $^{133}$Cs. The frequency of the end $\sigma_+$ transition of $^{87}$Rb overlaps the end $\sigma_-$ transition of $^{133}$Cs at 7.963 GHz with a DC field of 512 Gauss. The total continuously tunable frequency range was from 1.5 to 11.3 GHz when a DC field of up to 860 Gauss was applied.

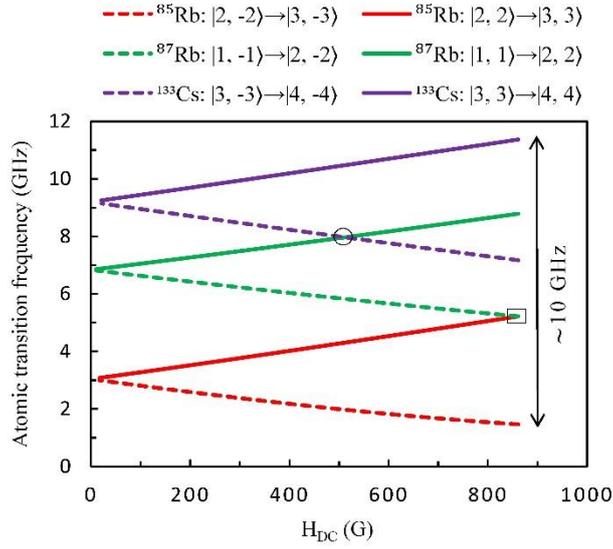

FIG. 2. Numerical calculation of the end Zeeman transition frequency as a function of the DC field applied for $^{85}$Rb (red), $^{87}$Rb (green), and $^{133}$Cs (purple). The dashed lines are end $\sigma_-$ transitions and the solid lines are end $\sigma_+$ transitions. The black circle and square represent the overlapping hyperfine transition frequencies between different species that correspond to the different DC fields.

An experimental setup for this frequency tunable MW magnetic field detection technique was developed as shown in Fig. 3. We used two laser sources in this experimental setup. The first laser source was a 1 MHz-linewidth distributed feedback (DFB) diode laser resonant on the $^{133}$Cs D2 line at 852 nm. The laser was frequency stabilized by the saturated absorption technique to the $F = 4 - F' = 4$ optical transition. The other laser source was a DFB locked onto the $F = 1 - F' = 2$ transition of the $^{87}$Rb D1 line at 795 nm also by a saturated absorption setup.

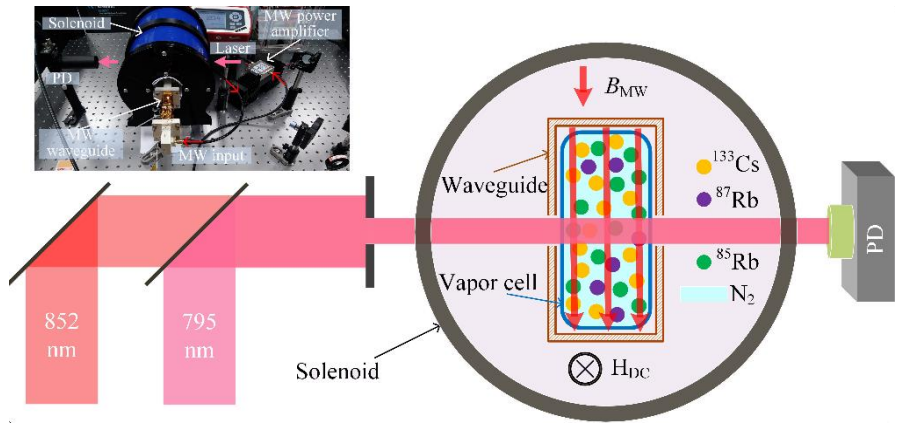

FIG. 3. Schematic of the experimental setup and photograph of the frequency tunable microwave (MW) magnetic field detection system based on Rabi resonance. Two laser beams were respectively locked onto the Cs D2 and Rb D1 lines by a saturated absorption technique. The hybrid cell contained $^{133}$Cs, $^{85}$Rb, $^{87}$Rb, and $N_2$. $B_{MW}$: microwave magnetic field; $H_{DC}$: DC magnetic field; PD: photodetector.

Both frequency-stabilized lasers were used to pump and probe atoms. The two laser beams were overlapped with a polarized beam splitter and then injected into a hybrid cell containing Cs and Rb atoms filled with $N_2$ buffer gas placed at the center inside a solenoid coil. The coil could provide a 525 Gauss DC magnetic field. A polarizer and a half wave plate (HWP) were placed ahead of the coil to ensure the laser beams were adequately linearly polarized. The diameter of the laser beams was 6 mm. The alkali atoms vapor cell was controlled to maintain a temperature of 35℃. The phase modulated MW field was generated by a commercial low noise MW signal synthesizer (E8257D, Keysight Technologies, Santa Rosa, CA, USA) and radiated by a wide band open-ended rectangular waveguide antenna. The MW field could excite the different magnetic dipole transitions of the Cs and Rb atoms depending on its frequency.

Figure 4 (a) reports the double resonance (DR) peaks for a frequency range from 5.79 to 10.51 GHz using our home-made solenoid coil. The signals were experimentally plotted with end $\sigma_-$ and $\sigma_+$ transitions of $^{87}$Rb (orange and purple) and $^{133}$Cs (green and red) under different DC fields. Only one laser beam was used when the data were collected, with the 795 nm laser on for the $\sigma_-$ and $\sigma_+$ transitions of $^{87}$Rb, and the 852 nm laser was on for $^{133}$Cs. The frequencies of the lasers were locked onto the auxiliary cells, and therefore their frequencies were detuned to resonance with the shifted hyperfine states, which would partially depopulate the state through optical pumping. Figure 4 (b) plots the DR signals of the $\sigma_-$ transition of $^{133}$Cs and $\sigma_+$ transition of $^{87}$Rb around the overlaps. When $H_{DC} = 512$ Gauss, the DR signal of the $\sigma_-$ transition of $^{133}$Cs was overlapped with the $\sigma_+$ transition of $^{87}$Rb. The frequency of the crossing point was 7.963 GHz. If the DC field $H_{DC}$ continued to increase, the frequency of Cs $\sigma_-$ transition shifted to a lower frequency than the $^{87}$Rb $\sigma_+$ transition. The experimental results shown in Figures 4 (a) and (b) show that the frequency tunable range of $^{87}$Rb and $^{133}$Cs atoms could be partially overlapped within a certain scope of the DC field, which indicates that the sensing frequency range of the MW magnetic field could be continuously tuned from 5.79 to 10.51 GHz. Ideally, the tunable frequency range would be further extended. The frequency limit was mainly due to technical limitations associated with the DC field or MW synthesizer.

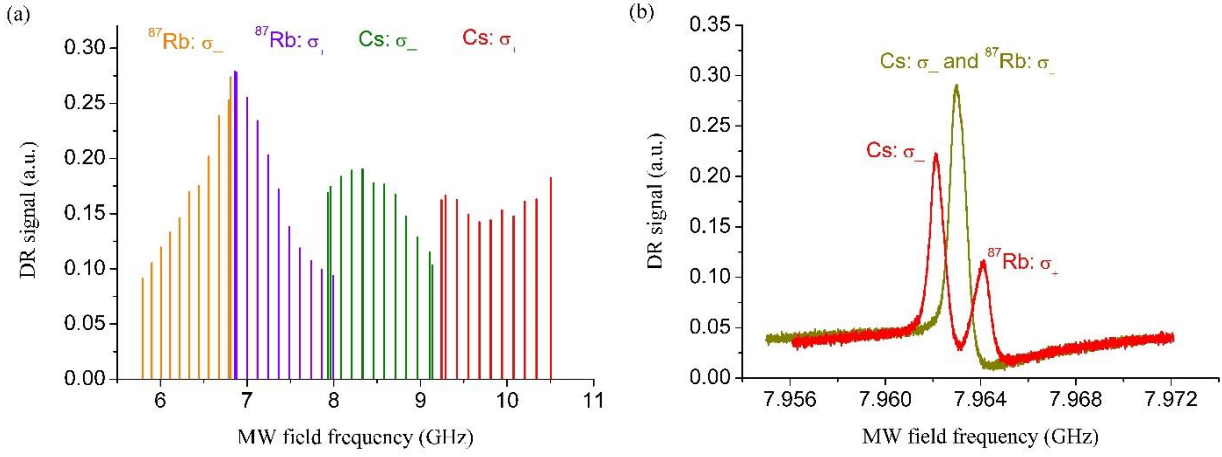

FIG. 4. (a) Double resonance (DR) signal of the frequency range from 5.79 to 10.51 GHz. Each peak corresponds to a DR signal of a different Zeeman transition when applying a different DC field. The four Zeeman transitions here are the end $\sigma_-$ (orange) and $\sigma_+$ (purple) transitions of $^{87}$Rb and the $\sigma_-$ (green) and $\sigma_+$ (red) transitions of $^{133}$Cs. (b) DR signals of the end $\sigma_+$ transition of $^{87}$Rb and the $\sigma_-$ transition of $^{133}$Cs around 7.963 GHz. The two DR signals were overlapped with a DC field of 512 Gauss. The frequencies of the two transitions continued to shift to the opposite direction as the DC field increased.

The sensing frequency of MW field detection is determined by the frequency of the DR signal. Figure 5 shows the full Rabi resonance signal lineshapes plotted at the sensing frequencies of other than 9.19 and 6.83 GHz when the DC field was applied. We used the end transitions of $^{133}$Cs and $^{87}$Rb to plot the Rabi resonance lineshapes in weak and strong DC fields, respectively. The measured Rabi frequency $\Omega$ could be derived from the phase modulation frequency, with $\omega_m$ corresponding to the peak amplitude of lineshape ($\Omega = 2\omega_m$). Figures 5 (a) and (b) show the Rabi resonance lineshape plotted at the sensing frequency of 9.201 and 6.122 GHz with a DC field of 3.5 and 350 Gauss, respectively. An unknow MW field was radiated to the hybrid alkali atoms vapor cell in free space. The first Rabi resonance lineshape was obtained with the end $\sigma_+$ transition of $^{133}$Cs, while the latter was obtained with the end $\sigma_-$ transition $^{87}$Rb. Both lineshapes were well fitted by the theoretical model Eq. (1).

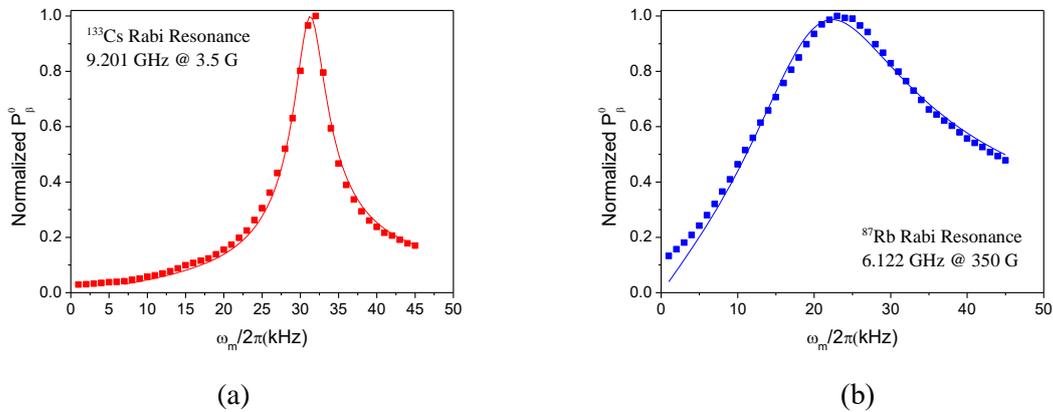

(a)      (b)

FIG. 5. Rabi resonance lineshapes at the sensing frequencies of 9.201 GHz (a) and 6.122 GHz (b). The driving DC fields were 3.5 and 350 Gauss, respectively. The end $\sigma_+$ transition of $^{133}$Cs and the end $\sigma_-$ transition of $^{87}$Rb were used. The signals were normalized.

To validate the MW detection capability of the Rabi resonance-based system at an extended sensing frequency, we measured different frequency MW magnetic fields, with a variety of MW powers. Figure 6 plots the measured Rabi frequency $\Omega = 2\omega_m$ as a function of $\sqrt{P_{in}}$ for MW magnetic fields of 5.79 GHz (red) and 10.51 GHz (purple), where $P_{in}$ is the MW incident power. As expected, the measured Rabi frequencies were linearly dependent on $\sqrt{P_{in}}$. Moreover, the slopes of the fits were approximately equal at all sensing frequencies. This confirms that the MW magnetic field detection technique based on Rabi resonance is applicable over a large sensing frequency range.

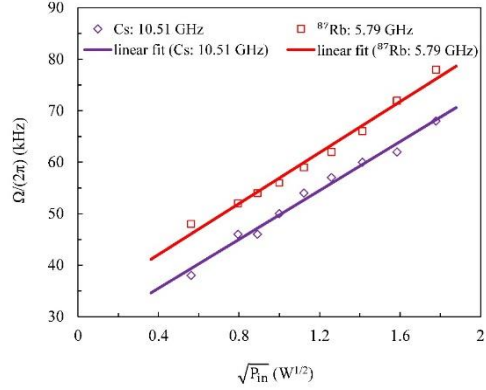

FIG.6. Measured Rabi frequency $\Omega = 2\omega_m$ as a function of $\sqrt{P_{in}}$ for microwave (MW) magnetic fields of 5.79 GHz (red) with $^{87}$Rb and 10.51 GHz (purple) with $^{133}$Cs.

Note that the Rb atoms in our hybrid vapor cell contained both $^{87}$Rb and $^{85}$Rb isotopes, meaning that our hybrid cell allows the simultaneous detection of the Rabi resonance of $^{85}$Rb. However, the open-ended waveguide antenna used in our experiment cannot work at the $^{85}$Rb frequency due to its frequency cutoff characteristics below 5.5 GHz. By using another horn antenna for the MW radiation, we successfully observed the Rabi resonance signal of $^{85}$Rb at 3.04 GHz, as shown in Fig. 7. Furthermore, as with $^{87}$Rb and $^{133}$Cs atoms, the measurement of Rabi frequency is quite straightforward. The Rabi resonance lineshape was fitted using Eq. (1). The measurement results indicated that the sensing frequency range of our MW detection technique could be extended to lower than 3 GHz, with an appropriate DC field and Zeeman transition.

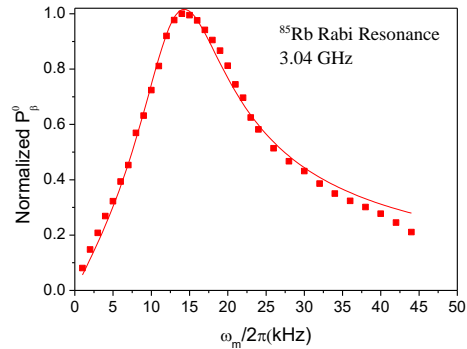

FIG. 7. Typical Rabi resonance lineshape for measuring a microwave (MW) field of 3.04 GHz. The signals were normalized.

To summarize, we have demonstrated a frequency tunable MW magnetic field detection technique based on Rabi resonance. We could finely control the sensing frequency of the MW magnetic field generated by applying a DC field and different Zeeman transitions with a hybrid vapor cell filled with Rb and Cs atoms. The experimental system consisted mainly of two DFB lasers locked onto the Rb D1 and Cs D2 lines, respectively, a coil that could generate a DC field up to 500 Gauss, and a hybrid vapor cell containing Rb and Cs atoms. The MW magnetic field detection was reported at 5.79 GHz with the $\sigma_-$ transition of $^{87}$Rb and at 10.51 GHz with the $\sigma_+$ transition of $^{133}$Cs. The Zeeman transition frequency of atoms could be shifted to even higher or lower values with a larger DC field. However, the DR signal and Rabi resonance signal of a certain Zeeman transition was decreased as the DC field increases, thus, the more species of atoms mixed in a single vapor cell, the larger the sensing frequency range with a relatively small DC field. This frequency tunable detection technique provides an important approach to the development of an atom-based and SI-traceable MW field sensor that could operate at any frequency.

This work was funded in part by the National Key Research and Development Program of China under Grant 2018YFF0212406, the Beijing Natural Science Foundation under Grant 4182078, and the National Natural Science Foundation of China under Grant 61601084.